\documentclass{sig-alternate}

\usepackage{amsmath}
\usepackage{graphicx}
\usepackage{mathtools}
\usepackage{tabularx}
\usepackage{rotating}
\usepackage{booktabs}
\usepackage{url}
\usepackage{color}

\usepackage{xspace}

\newdef{example}{Example}

\newcommand{\recomp}{\texttt{ReComp}\xspace}
\newcommand{\D}{\mathbf{D}}
\newcommand{\y}{\mathbf{y}}
\newcommand{\x}{\mathbf{x}}
\newcommand{\dt}{\mathit{dt}}

\newcommand{\exec}{\mathit{exec}}

\newcommand{\update}[2]{#2 \rightarrow #1}

\newcommand{\CV}{\mathit{CV}}
\newcommand{\OM}{\mathit{OM}}
\newcommand{\varst}{\mathit{varst}}

\newcommand{\prov}{\mathit{prov}}

\newcommand{\imphat}{\widehat{\mathit{imp}}}
\newcommand{\diff}[1]{\mathit{diff}_{#1}}
\newcommand{\diffhat}[1]{\widehat{\mathit{diff}}_{#1}}

\newcommand{\diffo}[1]{\mathit{diff}_{#1}}
\newcommand{\diffOM}{\mathit{diff}_{OM}}
\newcommand{\diffCV}{\mathit{diff}_{CV}}
\newcommand{\genes}{\mathit{genes}\ }

\begin{document}



\title{Preserving the value of large scale data analytics over time through selective re-computation
\titlenote{This work is supported by EPSRC grant EP/N01426X/1 (2016-2019) in the UK}}

\subtitle{\small Technical Report \\ School of Computing Science, Newcastle University \\ Oct. 2016}

\numberofauthors{3} 

\author{
\alignauthor
Paolo Missier\\
       \affaddr{School of Computing Science}\\
       \affaddr{Newcastle University, UK}\\
       \email{Paolo.Missier@ncl.ac.uk}
\alignauthor
Jacek Ca\l{}a\\
       \affaddr{School of Computing Science}\\
       \affaddr{Newcastle University, UK}\\
       \email{Jacek.Cala@ncl.ac.uk}
\alignauthor Manisha Rathi\\
       \affaddr{School of Computing Science}\\
       \affaddr{Newcastle University, UK}\\
       \email{Manisha.Rathi@ncl.ac.uk}
}

\maketitle
\begin{abstract}
A pervasive problem in Data Science is that the knowledge generated by possibly expensive analytics processes is subject to decay over time, as the data used to compute it drifts, the algorithms used in the processes are improved, and the external knowledge embodied by reference datasets used in the computation evolves.
Deciding when such knowledge outcomes should be refreshed, following a sequence of data change events, requires problem-specific functions to quantify their value and its decay over time, as well as models for estimating the cost of their re-computation.
What makes this problem challenging is the ambition to develop a decision support system for informing data analytics re-computation decisions over time, that is both 
generic and customisable.
With the help of a case study from genomics, in this vision paper we offer an initial formalisation of this problem, highlight research challenges, and outline a possible approach based on the collection and analysis of metadata from a history of past computations.
\end{abstract}

\section{Your data will not stay smart forever}

A general problem in Data Science is that the knowledge generated through large-scale data analytics tasks is subject to decay over time, following changes in both the underlying data used in their processing, and the evolution of the processes themselves.
In this paper we outline our vision for a general system, which we refer to as \recomp, that is able to make informed re-computation decisions in reaction to any of these changes.
We distinguish two complementary patterns, which we believe are representative of broad areas of data analytics.

\textbf{1. Forwards ReComp. } In this pattern, knowledge refresh decisions are triggered by changes that occur in the inputs to an analytics process, and are based on an assessment of the consequences of those changes on the current outcomes, in terms of expected value loss, or opportunities for value increase.

\textbf{2. Backwards ReComp.} Conversely, in this pattern the triggers are observations on the decay in the value of the outputs, and re-computation decisions are based on the expected value improvement following a refresh. 

In both cases, when a limited re-computation budget is available, estimates of the cost of refresh are needed. Cost may be expressed, for instance, as time and/or cost of cloud resource allocation.

To make these patterns concrete, we now present  one instance of each.

\subsection{Forwards: impact analysis}

Data-intensive workflows are becoming common in experimental science.
In genomics, for instance, it is becoming computationally feasible to process the human genome in search of mutations that may help diagnose a patient's genetic disease.
In this scenario, which we expand on in Sec.~\ref{sec:SVI}, a diagnosis given in the past may be affected by improvements in the underlying genome sequencing technology, but also possibly in the bioinformatics algorithms, and by updates in the external reference data resources like the many human variation databases~\cite{Missier2015,Cooper2011}.
In a Forwards ReComp scenario, each of these changes would trigger a decision process aimed at predicting which patients \textit{would benefit the most} from a reassessment of their diagnosis. A limited budget leads to a problem of prioritising re-computations over a subset of the patients' population, using estimates of the future cost of re-enacting the workflows.
A similar scenario occurs when long-running simulations are used e.g. to predict flood in large cities. In this case, the problem involves understanding the impact of changes to the urban topology and structure (new green areas, new buildings), without having to necessarily run the simulation anew every time.

\subsection{Backwards: cause analysis}

In machine learning, it is well-known that the predictive power of a trained supervised classifier tends to decay as the assumptions embodied by the data used for training are no longer valid.
When new ground truth observations become available while using the model, these provide a measure of actual predictive performance and of its changes over time, i.e., relative to the expected theoretical performance (typically estimated a priori using cross-validation on the training set). 
We may view the trained model as the ``knowledge outcome'' and the problem of deciding when to refresh (re-train) the model as an instance of Backwards ReComp.
Here the expected performance of the new model must be balanced against the cost of retraining, which is often dominated by the cost of generating a new training set.


\subsection{The ReComp Vision}  \label{sec:vision}

\begin{figure}[htbp]
  \centering
  \includegraphics[width=\linewidth]{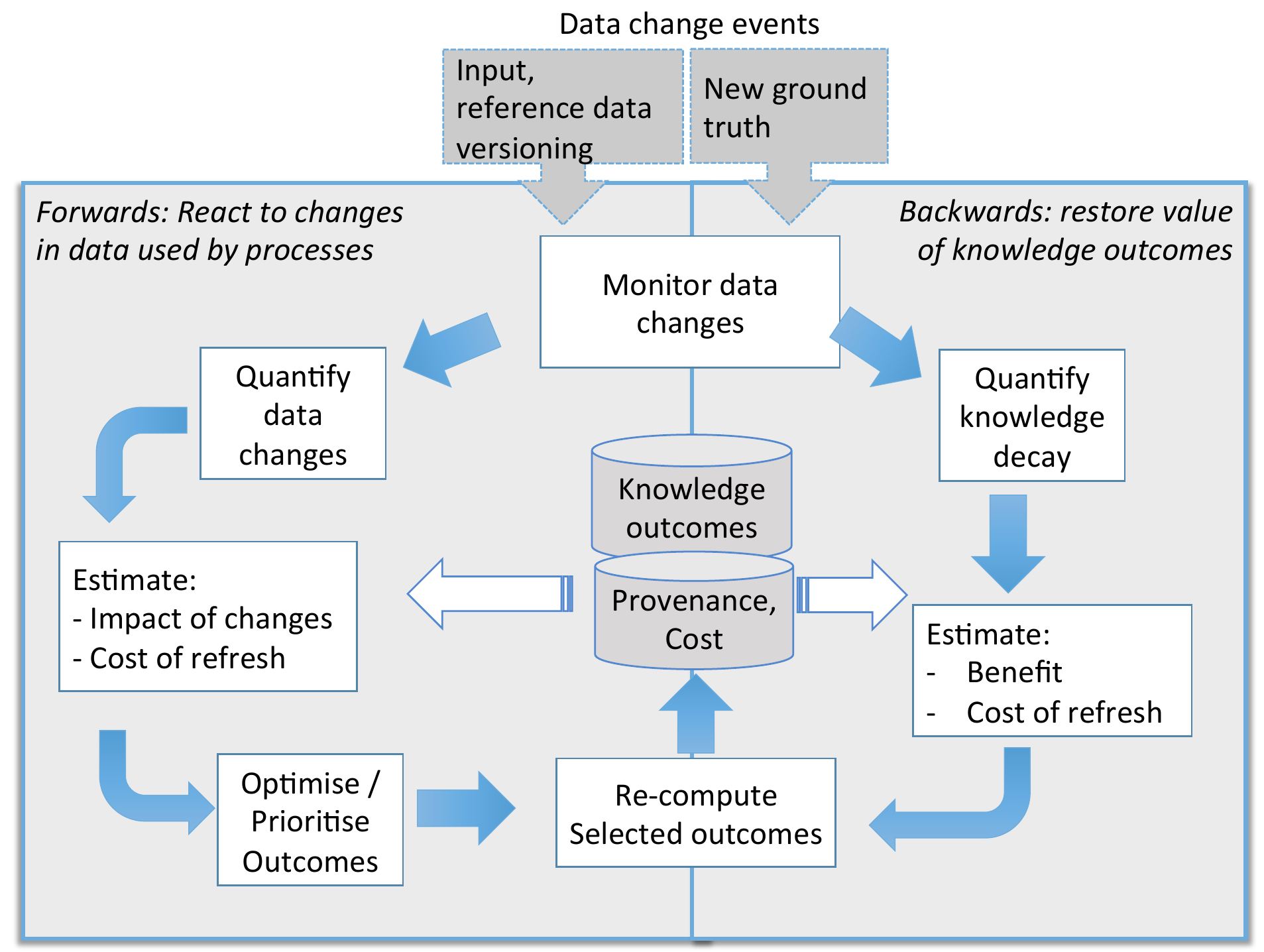}
	\caption{Reference ReComp patterns}
	\label{fig:recomp-vision}
\end{figure}

Fig. \ref{fig:recomp-vision} provides a summary of our vision of a \recomp meta-process for making recurring, selective re-computation decisions on a collection of underlying analytics processes for both these patterns.
In both cases, the meta-process is triggered by observations of data changes in the environment (top). 
In the Forwards pattern, on the left, these are new versions of data used by the process.
This pattern requires the ability to
(i) quantify the differences between two versions of a data,
(ii) estimate the impact of those changes on a process outcomes,
(iii) estimate the cost of re-computing a process,
(iv) use those estimates to select past process instances that optimise the re-computation effort subject to a budget constraint, and
(v) re-enact the selected process instances, entirely or partially.

The Backwards pattern, on the right, is triggered by changes in data that can be used to assess the loss of value of knowledge outcomes over time, such as new  ground truth data as mentioned earlier.
This pattern requires the ability to
(i) quantify the decay in the value of knowledge, expressed for instance in terms of model prediction power; 
(ii) estimate the cost and benefits of refreshing the outcome, and 
(iv) re-enact the associated processes.

Note that we only focus on changes in the data. 
For simplicity here we do not consider changes in the underlying processes, which are also relevant but require a separate formalisation, beyond the scope of this short paper.

To inform these decisions and realise these patterns, we envision a History database (centre). This contains both the outcomes that are subject to revision, and metadata about their provenance \cite{w3c-prov-overview} and their cost. 
Estimation models are learnt from the metadata, which is then updated following each re-computation cycle.

\subsection{Example: Genetic variants analysis}  \label{sec:SVI}
\begin{figure}
  \centering
  \includegraphics[width=\linewidth]{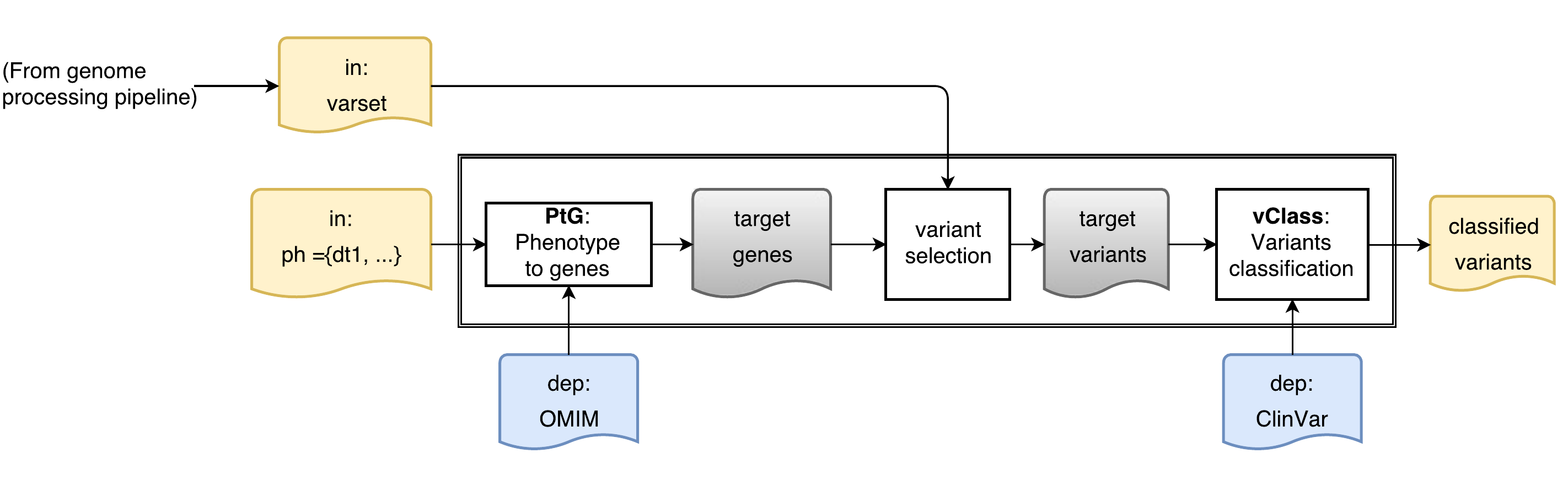}
  \caption{Sketch of the SVI workflow, with inputs $\x = [\mathit{varset}, \mathit{ph}]$ and external resources $\D = [\OM, \CV]$.}
  \label{fig:svi-pipeline}
\end{figure}

The Simple Variant Interpretation (SVI) process~\cite{Missier2015} is designed to support clinical diagnosis of genetic diseases. 
A patient's complement of \textit{variants}, or single-nucleotide gene mutations, is identified by processing the patient's genome. The process essentially identifies mutations by comparing the patient's to a reference genome. More precisely, the SVI workflow, sketched in Fig. \ref{fig:svi-pipeline}, takes a patient's variants (about 25,000) and a set of terms that describe the patient's \textit{phenotype},  and tries to establish the deleteriousness of the small subset of those variants that are relevant for the phenotype, by consulting external reference mutation databases.
In particular, SVI uses knowledge from the ClinVar\footnote{\url{www.ncbi.nlm.nih.gov/clinvar}} and OMIM Gene Map\footnote{\url{www.ncbi.nlm.nih.gov/omim}} reference databases, described in more detail later.

The reliability of the diagnosis depends upon the content of those databases. 
While the presence of deleterious variants may sometimes provide conclusive evidence in support of the disease hypothesis, the diagnosis is often not conclusive due to missing information about the variants, or due to insufficient knowledge in those databases.
As this knowledge evolves and these resources are updated, there are opportunities to revisit past inconclusive or potentially erroneous diagnoses, and thus to consider re-computation of the associated analysis. 
Furthermore, a patient's variants, used as input to SVI, may also be updated as sequencing and variant calling technology improve.

We use SVI in our initial experiments, as it is a small-scale but fair representative of large-scale genomics pipelines that also require periodic re-computation, such as those for variant calling that we studied in the recent past~\cite{Cala2015a}.

\subsection{Contributions}
We make the following contributions. (i) A formalisation of the selective re-computation problem, which due to space limitations is limited to the \textit{forwards} case, exemplified by the SVI case study; (ii) an outline of the associated research challenges,  and (iii) an initial analysis of the role of metadata, and specifically of provenance, as part of the \recomp vision.
 
This work reflects the initial phase of a project centred on the design of the \recomp meta-process (recomp.org.uk).
What makes \recomp particularly challenging is the ambition to develop a \textit{generic} and customisable decision support system for informing data analytics re-computation decisions over time, in a setting where most approaches appear to be problem-specific.

\section{Reacting to data change events} 

We formalise the \textit{forwards} pattern of the ReComp problem in more detail, assuming an ideal scenario where a history of past program executions has been recorded, each data item is versioned, and a family of \textit{data diff} functions, one for each of the data types involved in the computation, are available to quantify the extent of change between any two versions.

\subsection{Definitions} \label{sec:definitions}

\textbf{Executions.} 
Suppose we can observe a collection of $N$ executions of a analytics applications, which we represent as a single program $P$ for simplicity of exposition.
Each execution $i:1 \dots N$ takes input $x_i$ (possibly a vector of values) and may also use data queried from a set of \textit{reference datasets} $D = \{D_1 \ldots D_m\}$  during its execution, to produce a value $y_i$.
We assume that each of the $x_i$ and each $D_j \in D$ may have multiple versions, which are updated over time. 
We denote the version of $x_i$ at time $t$ as $x_i^t$, and the state of resource $D_j$ at $t$ as $d_j^t$.
For each execution, we also record its cost $c_i^t$ (this can be for example a time or monetary expression that summarises the cost of cloud resources).
We denote one execution of $P$ that takes place at time $t$ by:
\begin{equation}
\langle y_i^t, c_i^t \rangle = \exec(P, x_i^t, d^t) 
\label{eq:exec}
\end{equation}
where $d^t = \{ d_1^t \dots d_m^t\} $ is the state at time $t$ of each of the $D_j$.
As mentioned (Sec. \ref{sec:vision}), we assume that $P$ stays constant throughout.

\begin{example}
\sloppy SVI consists of one single process $P$, which initially is executed once for each new patient.
It takes an input pair $x = \langle \mathit{varset}, \mathit{ph} \rangle$ consisting of the set of that patient's variants and the patient's phenotype $\mathit{ph} = \{ \dt_1, \dt_2, \dots \}$ expressed using \textit{disease terms} $\dt_i$ from the OMIM vocabulary, for example \emph{Alzheimer's}.

SVI is a classifier that associates a class label to each input variant depending on their estimated deleteriousness, using a simple ``traffic light'' notation:
\[\y = \{ (v, \mathit{class}) | v \in  \mathit{varset}, \mathit{class} \in \{ \textsf{red}, \textsf{amber}, \textsf{green}\}\} \]
$D = \{\OM, \CV\}$  consists of two reference databases, OMIM GeneMap and Clinvar, which are subject to periodic revisions.
GeneMap maps human disorder terms $\dt$ to a set of genes that are known to be broadly involved in the disease:
\[\OM = \{ \langle \dt, \genes(\dt)  \rangle  \} \]
Similarly, ClinVar maintains a catalogue $V$ of variants, each located on a gene $g$, and 
it associates a status to each variant $v \in V$, denoted
$\varst(v) \in \{ \textsf{unknown}, \textsf{benign}, \textsf{pathogenic} \}$:
\[ \CV = \{ \langle v, g, \varst(v \rangle \}\]

SVI uses $\OM$ and $\CV$ to investigate a patient's disease, as shown in Fig. \ref{fig:svi-pipeline}.  
Firstly, the terms in $\mathit{ph}$ are used to determine the set of \textit{target genes} that are relevant for the disease hypothesis. 
These are defined as the union of all the genes in $genes(\dt) $ for each disease term $\dt \in  \mathit{ph}$.
%
Then, a variant $v \in \mathit{varset}$  is selected if it is located on the \textit{target genes}.
Finally, the selected variants are classified according to their labels from $\varst(v)$.
$\Box$
\end{example}

\textbf{Data version changes.} 
We write $\update{x_i^{t'}}{x_i^t}$ to denote that a new version of $x_i$ has become available at time $t'$,  replacing the version $x_i^t$ that was current at $t$.
Similarly, 
 $\update{d_j^{t'}}{d_j^t}$ denotes a new release of $D_j$ at time $t'$.
%

\textbf{Diff functions.} We further assume that a family of type-specific \textit{data diff} functions are defined that allow us to quantify the extent of changes.
Specifically:
\begin{align}
\diff{X}(x_i^t, x_i^{t'})  \quad \diff{Y}(y_i^t, y_i^{t'})
\label{eq:diff-xy}
\end{align}
compute the differences between two versions of $x_i$ of type $X$, and two versions of $y_i$ of type $Y$.
Similarly, for each source $D_j$, 
\begin{equation}
\diff{D_j}(d_j^t,d_j^{t'}) 
\label{eq:diff-d}
\end{equation}
quantifies the differences between two versions of $D_j$.
The values computed by each of these functions are going to be type-specific data structures, and will also depend on how changes are made available.
For instance, $d_j^t, d_j^{t'}$ may represent successive transactional updates to a relational database.
More realistically in our analytics setting, and on a longer time frame, these will be two releases of $D_j$, which occur periodically.
In both cases, $\diff{D_j}(d_j^t,d_j^{t'})$ will contain three sets of added, removed, or updated records, respectively.

\begin{example}
Considering that the set of terms $\dt$ in OMIM is fairly stable, $\diff{OM}(\OM^t, \OM^{t'})$ returns updates in their mappings to genes that have changed between the two versions (including possibly new mappings):
\begin{align*}
 &\diffOM(\OM^t, \OM^{t'}) = \\
& \{\langle  t, genes(\dt) \rangle | genes(\dt) \neq genes'(\dt) \} 
\end{align*}
where $genes'(\dt)$ is the new mapping for $\dt$ in $\OM^{t'}$.

The difference between two versions of ClinVar consists of three sets: new, removed, and status-changed variants:
\begin{align*}
\diffCV&(\CV^t, \CV^{t'}) = \\
&\{ \langle v, \varst(v)  | \varst(v) \neq \varst'(v) \} \\
& \cup \CV^{t'} \setminus \CV^t \cup \CV^t \setminus \CV^{t'}
\label{eq:diff-cv}
\end{align*}
where $\varst'(v)$ is the new class associate to $v$ in $\CV^{t'}$.
$\Box$
\end{example}

%
\textbf{Change Impact.} To describe the \textit{impact} of a single change that occurs at time $t'$ on an output $y_i^t$ that is current at some $t < t'$,
suppose we have computed the new $y_i^{t'}$ using the new version of the data. 
For instance, if the change is $\update{d_j^{t'}}{d_j^t}$, we would have computed:
\begin{equation}
\langle y_i^{t'}, c_i^{t'} \rangle = \exec(P, x_i^{t'}, d^{t'}) 
\label{eq:refreshed-exec}
\end{equation}
where $d^{t'} = \{ d_1^t \dots d_i^{t'} \dots d_m^t \}$.
We define the impact of this change using a type-specific function $f_Y()$ with range normalised to $[0,1]$, defined on the difference between the two versions of $y_i$:
\begin{equation}
\mathit{imp}(\update{d_j^{t'}}{d_j^t}, y_i^t) = f_Y(\diff{Y}(y_i^t, y_i^{t'}))  \in [0,1]
\label{eq:imp}
\end{equation}
	where $y_i^{t'}$ is computed as in (\ref{eq:refreshed-exec}).
\sloppy In the case of our classified variants, for instance, $f_Y()$ could be defined as $f_Y(\diff{Y}(y_i^t, y_i^{t'})) = 0$ if the diagnosis has not changed between two versions, and 1 if it has changed. 

\subsection{Problem statement} 
Suppose a change is detected at $t'$, for simplicity let it be $\update{d_j^{t'}}{d_j^t}$ as above.
Let $O^t = \{ y_1^t, \dots y_N^t\}$ denote the set of all outcomes that are current at time $t$.

The ReComp goal is to select the optimal subset $O_{rc}^t \subseteq O^t$ of outcomes that would maximise the overall impact of the change if they were re-computed, subject to a budget $C$:
\begin{equation}
\max_{O_{rc}^t \subset O^t} \sum_{y_i \in O_{rc}^t}\mathit{imp}(\update{d_j^{t'}}{d_j^t}, y_i^t) \text{,} \quad  \sum_{i:1}^{N} c_i^{t'} \leq C
\end{equation}
As neither the impact nor the actual re-computation costs are known, however, solving the problem requires first that we learn a set of cost and impact estimators for them:
\begin{equation}
\{ \langle \imphat(\update{d_j^{t'}}{d_j^t}, y_i^t),  \hat{c}_i^{t'}  \rangle | y_i^t \in O^t\}
\label{eq:imp-est}
\end{equation}

The optimisation problem can thus be written as:
\begin{align}
\max_{O_{rc}^t \subset O^t} \sum_{y_i \in O_{rc}^t}\imphat(\update{d_j^{t'}}{d_j^t}, y_i^t) \textbf{,} \quad \sum_{i:1}^{N} \hat{c}_i^{t'} \leq C
\label{eq:recomp-estimators}
\end{align}

\section{ReComp Challenges}  \label{sec:challenges}
A number of process and management challenges underpin this optimisation goal for the Forwards ReComp pattern.

\subsection{Process Management Challenges}

\textbf{1. Optimisation of re-computation effort.}
Firstly, note that we must solve one instance of (\ref{eq:recomp-estimators}) for each data change event.
Each of those instances can be formulated as a 0-1 knapsack problem in which we want to find vector
$\mathbf{a} = [a_1 \dots a_n] \in \{0,1\}^N$ that achieves
\begin{equation}
\max \sum_{i:1}^{N} v_i a_i \text{ subject to }  \qquad \sum_{i:1}^{N} w_i a_i \leq C 
\label{eq:knapsack}
\end{equation}
where $v_i = \imphat(\update{d_j^{t'}}{d_j^t}, y_i^t)$, $w_i = \hat{c}_i^{t'}$.\\

A further issue is whether multiple changes, i.e., to different data sources, should be considered together or separately. 
Also, in some cases it may be beneficial to group multiple changes to one resource, i.e., given  $\update{d_j^{t'}}{d_j^t}$, we may react immediately, or rather wait for the next change $\update{d_j^{t"}}{d_j^{t'}}$ and react to $\update{d^{t''}}{d^t}$.
%
%
%

\textbf{2. Partial recomputation.}
When $P$ is a \textit{black box} process, it can only be re-executed entirely from the start.
However, a \textit{white-box} $P$ such as a workflow, as in the case of SVI, may benefit from known techniques for ``smart re-run'', such as those developed in the context of scientific data processing using workflow management systems ~\cite{Altintas2006, Ludascher2006}.
%
Specifically, suppose that a granular description of $P$ is available, in terms of a set of processing blocks $\{ P_1 \dots P_l \}$ where in particular some $P_j$ encodes a query to $D_j$. 
These, along with dataflow dependencies of the form: $P_i \rightarrow P_j$, form a directed workflow graph. 

If re-computation of $P$ is deemed appropriate following a change in $D_j$, logically there is no need to restart the computation from the beginning, as long as it includes $P_j$ (because we know that a new execution of $P_j$ will return an updated result).
In theory, the exact minimal subgraph of $P$ that must be recomputed is determined by the available persisted intermediate data, saved during prior computations~\cite{Ludascher2006}. 
An architecture for realising this idea is also presented in \cite{Koop2010}.
In practice, however, for data analytics tasks where intermediate data often outgrow the actual inputs by orders of magnitude, the cost of persisting all intermediate results may be prohibitive. An open problem, partially addressed in \cite{Woodman2015}, is therefore to balance the choice of intermediate data to retain in view of a potential future re-computation, with its cost.


\textbf{3. Learning cost estimators.}
This problem has been addressed in the recent past, but mainly for specific scenarios that are relevant to data analytics, namely workflow-based programming on clouds and grid, \cite{Pietri2014,Malik2013}. But for instance \cite{Miu2012} showed that runtime, especially in the case of machine learning algorithms, may depend on features that are specific to the input, and thus not easy to learn.

\textbf{4. Process reproducibility issues.}
Actual re-computation of older processes $P$ may not be straightforward, as it may require redeploying $P$ on a new infrastructure and ensuring that the system and software dependencies are maintained correctly, or that the results obtained using new versions of third party libraries remain valid.
Addressing these architectural issues is a research area of growing interest \cite{freire_et_al:DR:2016:5817,burgess2016alan,Stodden2014}, but not a completely solved problem.

\subsection{Data Management Challenges}

\textbf{5. Learning impact estimators.}
Addressing the optimisation problem (\ref{eq:recomp-estimators}) requires that we first learn impact estimators  (\ref{eq:imp-est}).
In turn, this requires first estimating the differences $\diffhat{Y}(y_i^t, y_i^{t'})$ for any $y_i^t \in O^t$ and for any data change, where the estimators are going to be data- and change-specific and thus, once again, difficult to generalise.
This is a hard problem, as in particular it involves estimating the difference  $\diffo{Y}(y,y')$ between two values $y = f(x_1 \dots x_k)$, $y' = f(x'_1 \dots x'_k)$ for an unknown function $f$, given changes to some of the $x_i$ and the corresponding $\diffo{X}(x_i,x_i')$.
Clearly, some knowledge of function $f_Y()$ is required, which is also process-specific and thus difficult to generalise into a reusable re-computation framework.

\begin{example}
\sloppy Recalling our example binary impact function $f_Y()$ for $\CV$, we would like to predict whether any new variant added to $\CV^{t'}$ will change a patient's diagnosis.
Using forms of provenance, some of which is described later (Sec.\ref{sec:provenance}), 
we may hope not only to determine whether the variant is relevant for the patient, but also whether the new variant will change the diagnosis or it will merely reinforce it. 
This requires domain-specific rules, however, including checking whether other benign/deleterious variants are already known, and checking the status of an updated or new variant.
$\Box$
\end{example}

\textbf{6. Proliferation of specialised \textit{diff} functions.}
Suppose processes $P_1$ and $P_2$ retrieve different attributes from the same relational database $D_j$. Clearly, for each of them only changes to those attributes matter.
Thus, data diff functions such as those defined in Sec.~\ref{sec:definitions} are not only type-specific but also query-specific. 
For $K$ processes and $M$ resources, this potentially leads to the need for $KM$ specialised \textit{diff} functions.

%

%

\textbf{7. Managing data changes.}
There are practical problems in managing multiple versions of large datasets, each of which may need to be preserved over time for potential future use by \recomp.
Firstly, each resource will expose a different version release mechanism, standard version being the simple and lucky case.
Once again, observing changes in data requires source-specific solutions.
Secondly, the volume of data to be stored, multiplied by all the versions that might be needed for future re-computation, leads to prohibitively large storage requirements.
Thus, providers' limitations in the versions they make available translates into a challenge for \recomp.



\textbf{8. Metadata formats.}
\recomp needs to collect and store two main types of metadata, the detailed cost of past computations of $P$, which form ground truth data from which cost estimators can be learnt; and provenance metadata, as discussed next (Sec. \ref{sec:provenance}).
The former is a simpler problem, requiring the definition of a new format which, to the best of our knowledge, does not currently exist.
Provenance, on the other hand, has been recorded using a number of formats, which are system-specific.
Even when the PROV provenance model \cite{w3c-prov-dm} is adopted, it can be used in different ways despite being designed to encourage interoperability.
Our recent study~\cite{Oliveira2016} shows that the ProvONE extension to PROV (\url{https://purl.dataone.org/provone-v1-dev}) is a step forward to collect interoperable provenance traces, but it still limited in that it assumes that the traced processes are similar and implemented as a workflow.

\subsection{The ReComp meta-process}

To address these challenges, we have recently started to design a meta-process that can \textbf{observe executions} of the form (\ref{eq:exec}), \textbf{detect and quantify data changes} using \textit{diff} functions (\ref{eq:diff-xy}, \ref{eq:diff-d}), and \textbf{control re-computations} (\ref{eq:refreshed-exec}).

\recomp is an exercise in metadata collection and analysis. 
As suggested in Fig.\ref{fig:recomp-vision}, it relies on a history database that records details of all the elements that participate in each execution, as well as on the provenance of each output $y_i$, to provide the ground data from which estimators can hopefully be learnt.

However, not all processes and runtime environments are \textit{transparent} to observers, i.e., they may not allow for detailed collection of cost and provenance metadata.
Thus, we make an initial broad distinction between \textit{white-box} and \textit{black-box} \recomp, depending on the level of detail at which past computations can be observed, and the amount of control we have on performing partial or full re-computations on demand.

\section{Provenance in white-box Recomp}  \label{sec:provenance}
As an example of the role of metadata, we analyse how provenance might be used in a \textit{white-box}, fully transparent \recomp setting. 
Our goal is to reduce the size of the optimisation problem, that is, to identify those $y^t \in O^t$ that are \textit{out of scope} relative to a certain data change: these are the outputs that are definitely \textit{not} going to be affected by the change, and can therefore be ignored.
Formally, we want to determine the outputs $y_i^t \in O^t$ such that for a change $\update{d_j^{t'}}{d_j^t}$, we can determine that 
\[\mathit{imp}(\update{d_j^{t'}}{d_j^t}, y_i^t) = 0\]

For example, the scope of a change in ClinVar that reflects a newly discovered pathogenic status of a variant can be safely restricted to the set of patients who exhibit that mutation in one of the genes that are associated with their phenotype.

To achieve this filtering in a generic way, suppose we have access to the provenance of each $y_i^t$.
While this term refers generally to the history of data derivations from inputs to outputs through the steps of a process \cite{w3c-prov-overview}, in this setting we are only interested in recording which data items from $D_j$ were used by $P$ during execution.
In a particularly favourable but also common scenario, suppose that $D_j$ consists of a set of records, and that $P$ interacts with $D_j$ through well defined queries, denoted $Q_{D_j}$, using for instance a SQL or a keyword search interface.
Ordinarily, the provenance of $y_i^t$ would include all the data returned by execution of each of those queries: $Q_{d_j^t}$, along with the derivation relationships (possibly indirect) from those to $y_i^t$.
Instead, here we take an \textit{intensional} approach and record the queries themselves as part of the provenance:
\[\prov(y_i^t) = \{Q_{D_j}, j:i \dots m\}\]
where  each query is specific to the execution that computed $y_i^t$.
The rationale for this is that, by definition, an output $y_i^t$ is in the scope of a change to $d_j$ if and only if $P$ used any of the records in  $\diff{D_j}(d_j^t, d_j^{t'})$, that is, if and only if $Q_{D_j}$ returns a non-empty result when executed on the \textit{difference} $\diff{D_j}(d_j^t, d_j^{t'})$.

In practice, when $D_j$ is a set of records, we may naturally also describe $\diff{D_j}(d_j^t, d_j^{t'})$ as comprising of three sets of records $r$: new:$r \in d_j^{t'} \setminus d_j^{t}$, removed: $r \in d_j^{t} \setminus d_j^{t'}$, and updated: $r \in d_j^{t'} \cap d_j^{t}$ where some value has changed. 
This makes querying the differences a realistic goal, requiring minor adjustments to  $Q_{D_j}$ (to account for differences in format), i.e., we can assume we can execute $Q_{D_j}(d_j^{t'} \setminus d_j^{t})$, $Q_{D_j}(d_j^{t} \setminus d_j^{t'})$, and $Q_{D_j}(d_j^{t'} \cap d_j^{t})$.

\begin{example}
Consider patient Alice, whose phenotype is simply \emph{Alzheimer's}.
For SVI, this is also the keyword query to GeneMap: $Q_{\OM}$ = ``Alzheimer's''.
Suppose that performing the query at time $t$ returns just one gene: $Q_{\OM}(om^t) = \{ \texttt{PSEN2}\}$.
SVI then uses that gene to query $\CV$, and suppose that nothing is known about the variants on this gene:
$Q_{\CV}(cv^t) = \emptyset$.
At this point, the provenance of SVI's execution for Alice consists of the queries:
$\{Q_{\OM} \equiv  ``Alzheimer's'', Q_{\CV} \equiv ``\texttt{PSEN2}''\}$.

Suppose that at a later time $t'$ $\CV$ is updated to include just one new deleterious variant along with the gene it is situated on: $\langle \texttt{227083249}, \texttt{PSEN2}, \texttt{pathogenic} \rangle$.
When we compute $\diff{\CV}(cv^{t}, cv^{t'})$, this tuple is  included in $cv^{t'} \setminus cv^t$ and is therefore returned by a new query $Q_{\CV}$ on this differerence set, indicating that Alice is in the scope of the change.
 In contrast, executing on the same diff set a similar $\CV$ query from another patient's provenance, where \texttt{PSEN2} is not a parameter, returns the empty set, signalling that the patient is definitely not affected by the change.
 $\Box$
\end{example}

Note that a similar idea, namely of exploiting provenance records for partial re-computation, has been studied before in the Panda system \cite{Ikeda2011,Ikeda2010}, with the goal to determine precisely the fragment of a data-intensive program that needs to be re-executed in order to \textit{refresh} stale results.
However, its applicability requires full knowledge of the specific queries, which is not required here.
A formal definition of correctness and minimality of a provenance trace with respect to a data-oriented workflow is also proposed by members of the same group \cite{Ikeda2013}. 
The notion of \textit{logical provenance} that follows may be useful in our context, too, once it is mapped to the PROV data model \cite{w3c-prov-dm} that has since emerged as a standard for representing provenance.

Note also, that the technique just sketched will only go as far as narrowing the scope of a change, but will reveal little about its impact.
Still, in some cases we may be able to formulate simple domain-specific rules for qualitative impact estimation, that reflect our propensity to accept or prevent false negatives, i.e., ignoring a change that will have an impact. An example of such a conservative rule would be  ``if the change involves a new \textit{deleterious} variant, then re-compute all patients who are in the scope for that change''.

The example given earlier illustrates how queries saved from previous executions can be used to determine the scope of a change, assuming implicitly that the queries themselves remain constant. 
However this assumption can be easily violated, including in our running example. 
Suppose that at time $t$ $\OM$ is updated instead of $\CV$, for instance with the knowledge that an additional gene \texttt{X} is now known to be implicated in Alzheimer's'. 
We now have $Q_{\OM}(om^{t'}) = \{ \texttt{PSEN2}, \texttt{X}\}$, therefore $Q_{\CV} \equiv ``\texttt{PSEN2},  \texttt{X}''$ rather than just ``\texttt{PSEN2}'' as recorded in the provenance.
This brings the additional complication that queries stored in the provenance may need to be updated prior to being re-executed on the diff records.

Finally, note that in this specific example, when the change occurs in the input, that is, in the patient's genome, the scope of the change consists of just one patient.
In this case, it may well be beneficial to always re-compute, as computing $\diff{X}(x_i^t, x_i^{t'})$ to determine which parts of the genome have changed and whether the change will have any impact may be just as expensive, and thus inefficient. These questions are the subject of our current experimentation.

\subsection{Conclusions}
In this paper we have made the case for a new strand of research to investigate strategies for the selective, recurring re-computation of data-intensive analytics processes when the knowledge they generate is liable to decay over time.
Two complementary patterns are relevant in this setting, i.e. \textit{forwards impact analysis}, and \textit{backwards cause analysis}.
With the help of a case study in genomics we offered a simple formalisation of the former,\footnote{Analysis of the specific ``backwards'' cases will appear in a separate contribution.} and outlined a number of challenges, which arise when one sets out to design a generic and reusable framework for a broad family of underlying analytics processes.

To begin addressing these problems, we propose a \recomp meta-process that is able to collect metadata (cost, provenance) on a history of past computation and use it to learn cost and impact estimators, as well as to drive partial re-computation on a subset of prior outcomes.
As an example of our early investigation in this direction, we have discussed the role of data provenance in an ideal ``white box'' scenario.

\end{document}